\documentclass[prl,twocolumn,preprintnumbers,amsmath,amssymb,mathabx,textcomp,showpacs,superscriptaddress,lengthcheck]{revtex4}

\usepackage{graphicx}
\usepackage{dcolumn}
\usepackage{bm}
\usepackage{wasysym}

\newcommand{\halff}{\textstyle\frac{1}{2}}

\newcommand{\compressed}{{\textit{c}}}
\newcommand{\uncompressed}{{\textit{u}}}
\newcommand{\low}{\compressed}
\newcommand{\up}{\uncompressed}

\newcommand{\totalTime}{T}
\newcommand{\residenceTime}{t_\compressed}
\newcommand{\switchTime}{\tau_\compressed}

\newcommand{\tauDiffusiveStep}{\tau_D}

\newcommand{\Llobe}{L_1}

\newcommand{\ConfigurationVolume}{\Omega}
\newcommand{\OmegaLow}{\ConfigurationVolume_\low}
\newcommand{\OmegaUp}{\ConfigurationVolume_\up}
\newcommand{\OmegaTot}{\ConfigurationVolume}
\newcommand{\OmegaHardRods}{\omega}

\newcommand{\freeSpace}{\Delta}
\newcommand{\freeSpaceLow}{\freeSpace_\low}
\newcommand{\freeSpaceUp}{\freeSpace_\up}

\newcommand{\DlongTime}{D}

\newcommand{\numberOfJumps}{n}

\begin{document}
\title{A minimal model for kinetic arrest}
\author{P. Pal}
\altaffiliation{also Department of Applied Physics}
\affiliation{Department of Mechanical Engineering, Yale University, New Haven, CT 06520-8286.}
\author{C. S. O'Hern}
\altaffiliation{also Department of Physics}
\affiliation{Department of Mechanical Engineering, Yale University, New Haven, CT 06520-8286.}
\author{J. Blawzdziewicz}
\altaffiliation{also Department of Physics}
\affiliation{Department of Mechanical Engineering, Yale University, New Haven, CT 06520-8286.}
\author{E. R. Dufresne}
\altaffiliation{also Departments of Physics and  Chemical Engineering}
\affiliation{Department of Mechanical Engineering, Yale University, New Haven, CT 06520-8286.}
\author{R. Stinchcombe}
\affiliation{Rudolf Peierls Centre for Theoretical Physics, University of Oxford, 
1 Keble Road, Oxford, OX1 3NP, United Kingdom.}

\begin{abstract}
To elucidate slow dynamics in glassy materials, we introduce the {\it
Figure-8 model} in which $N$ hard blocks undergo Brownian motion
around a circuit in the shape of a figure-8.  This system
undergoes kinetic arrest at a critical packing fraction $\phi=\phi_g <
1$, and for $\phi\approx\phi_g$ long-time diffusion is controlled
by rare, cooperative `junction-crossing' particle rearrangements.  We
find that the average time between junction crossings $\tau_{JC}$, and
hence the structural relaxation time, does not simply scale with the
configurational volume $\OmegaLow$ of transition states, because
$\tau_{JC}$ also depends on the time to complete a 
junction crossing.  The importance of these results in understanding
cage-breaking dynamics in glassy systems is discussed.

\end{abstract}
\date{\today}
\maketitle

Glass transitions occur in myriad systems that span a wide range of
lengthscales \cite{angell} including atomic, polymeric, and colloidal
systems.  When cooled or compressed sufficiently fast, glass-forming
materials undergo a transition from an ergodic liquid state to an
amorphous solid-like glassy state \cite{pusey,blaaderen,weeks}.  Glassy
dynamics is characterized by several common features
\cite{debenedetti}.  For example, the viscosity and structural
relaxation times diverge super-Arrheniusly near the glass transition,
and the long-time self-diffusion constant becomes extremely small.
Correspondingly, a plateau develops in the particle mean-square
displacement (MSD).  As the system approaches the glass transition,
the plateau extends to longer and longer times
\cite{doliwa,thirumalai,miyagawa}, signaling kinetic arrest associated
with the formation of cages of neighbors around each particle.
Structural relaxation occurs through a series of rare cage-breaking
events, in which particles in the nearest-neighbor and further shells
move cooperatively so that caged particles can escape \cite{weeks}.

\begin{figure}
\begin{center}
\begin{tabular}{cccc}
\mbox{\scalebox{0.35}{\includegraphics{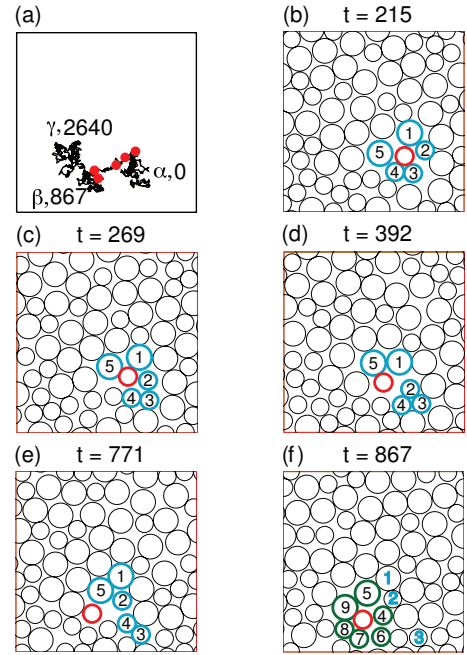}}}
\end{tabular}
\end{center}
\vspace{-0.2in}
\caption{\label{cage} (a) Particle trajectory taken from molecular
dynamics simulations of $64$ bidisperse hard disks 
over a time period in which the focus particle explores three cages
$\alpha$, $\beta$, and $\gamma$.  The cage-entrance times are 
provided.  The filled red circles correspond to
the snapshots in (b)-(f), which are labeled by time. The focus
particle is outlined in red and particles forming the $\alpha$
($\beta$) cage are outlined in blue (green).}
\vspace{-0.25in}
\end{figure}

The last feature is emphasized in Fig.~\ref{cage}, where we show
results from simulations of a 50\%--50\% bidisperse mixture of hard
disks with diameter ratio $1.4$ in the supercooled liquid regime.  In
Fig.~\ref{cage} (a), we plot displacements of a focus particle as it
moves between three cages $\alpha$, $\beta$, and $\gamma$.  In
Fig.~\ref{cage} (b)-(f), we monitor the focus particle and its
neighbors as it breaks out of cage $\alpha$ and becomes trapped in
$\beta$.  Two processes are required for cage-breaking to
occur. First, an opening must appear in the shell of nearest neighbors
(cf.\ panels (c) and (d)) and second, particles beyond the
nearest-neighbor shell must make sufficient room to accommodate the
escaping particle (cf. panels (d) and (e)).  If the material beyond
the nearest-neighbor shell behaves as a fluid, the second process
occurs easily, and the first is the rate controlling step.  However,
at higher packing fractions $\phi$ structural relaxation outside the
nearest-neighbor shell relies on other cage rearrangements.  As the
system approaches the glass transition, particles cannot escape from
their cages, because the surrounding particles, trapped in their own
cages, are unable to make room.  Geometrically, this is similar to the
mechanism responsible for rush-hour traffic jams, where cars cannot
exit an intersection because there is not enough room in the street in
front of them.  This, in turn, prevents cars in the perpendicular
direction from entering the intersection, which causes a cascade of
delays, leading to a city-wide traffic jam.

\begin{figure}
\begin{center}
\begin{tabular}{cccc}
\mbox{\scalebox{0.4}{\includegraphics{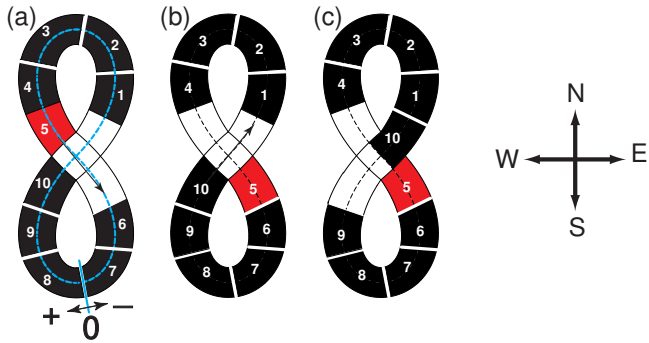}}}
\end{tabular}
\end{center}
\vspace{-0.3in}
\caption{\label{Figure-8Geometry}Schematic of the Figure-8 model near
$\phi_g$. (a)-(c) depict a junction-crossing event.  Initially half of
the particles are in each lobe.  Between (a) and (b) particle $5$
crosses the junction. Between (b) and (c) particle $10$ moves into the
junction in the direction of the arrow. In (c) the center of particle
$10$ resides in the upper lobe, which completes the junction-crossing
event.}
\vspace{-0.2in}
\end{figure}

To elucidate geometrical mechanisms responsible for extremely slow
dynamics in glassy materials, we introduce a `minimal' {\it Figure-8
model} that captures fundamental aspects of caging and cooperative
motion.  As illustrated in Fig.~\ref{Figure-8Geometry}, in our model
$N$ hard blocks undergo single-file diffusion around a continuous
course in the shape of a figure-8.  Kinetic arrest at large $\phi$
occurs because particles moving in one direction must vacate the
junction to allow those moving in the perpendicular direction to pass
through the junction.  The Figure-8 model has several appealing
features.  First, it is one of the simplest continuum models that
captures kinetic arrest (and the dependence of the structural
relaxation time on $\phi$ can thus be calculated analytically).
Second, to mimic a glassy material, the Figure-8 model can be
generalized to the `Manhattan' model, which includes an arbitrary
number of intersections and particles per lobe.  Third, experimental
realizations of the Figure-8 model can be performed, for example, by
confining colloidal suspensions in narrow channels
\cite{lutz,lin}. The Figure-8 model may provide insights into
important unanswered questions concerning glassy systems: 1) What
mechanisms give rise to cage-breaking events and how are they related
to dynamical heterogeneities \cite{string,ediger2}?  2) Why does
significant slowing down occur in dense particulate systems below
random close packing? 3) What is the form of the divergence of the
structural relaxation time near the glass transition?

We now provide a detailed description of the Figure-8 model.  The
upper and lower lobes of a channel of length $L$ intersect at a square
junction with unit dimensions.  The particles (hard blocks with
length $l=1$) undergo single-file Brownian diffusion,
implemented numerically using Monte-Carlo single-particle moves chosen
from a Gaussian distribution.  The total contour length $L$ and
average gap $\Delta$ per particle satisfy the relation $\phi = N/L =
1/(1+\Delta)$.  The particles are not allowed to turn at the
intersection.  Therefore, particles move through the junction in one
of two possible modes: (1) from north west to south east and vice
versa or (2) from north east to south west and vice versa.  To switch
modes, particles moving in one mode must vacate the junction to allow
particles in the other mode to enter the junction.  We focus our
analysis on small systems with $2\le N\le20$ particles to imitate
cooperative cage rearrangements in finite-size local regions in
glass-forming liquids (such as those depicted in Fig.~\ref{cage}).

\begin{figure}
\begin{center}
\begin{tabular}{cccc}
\mbox{\scalebox{0.12}{\includegraphics{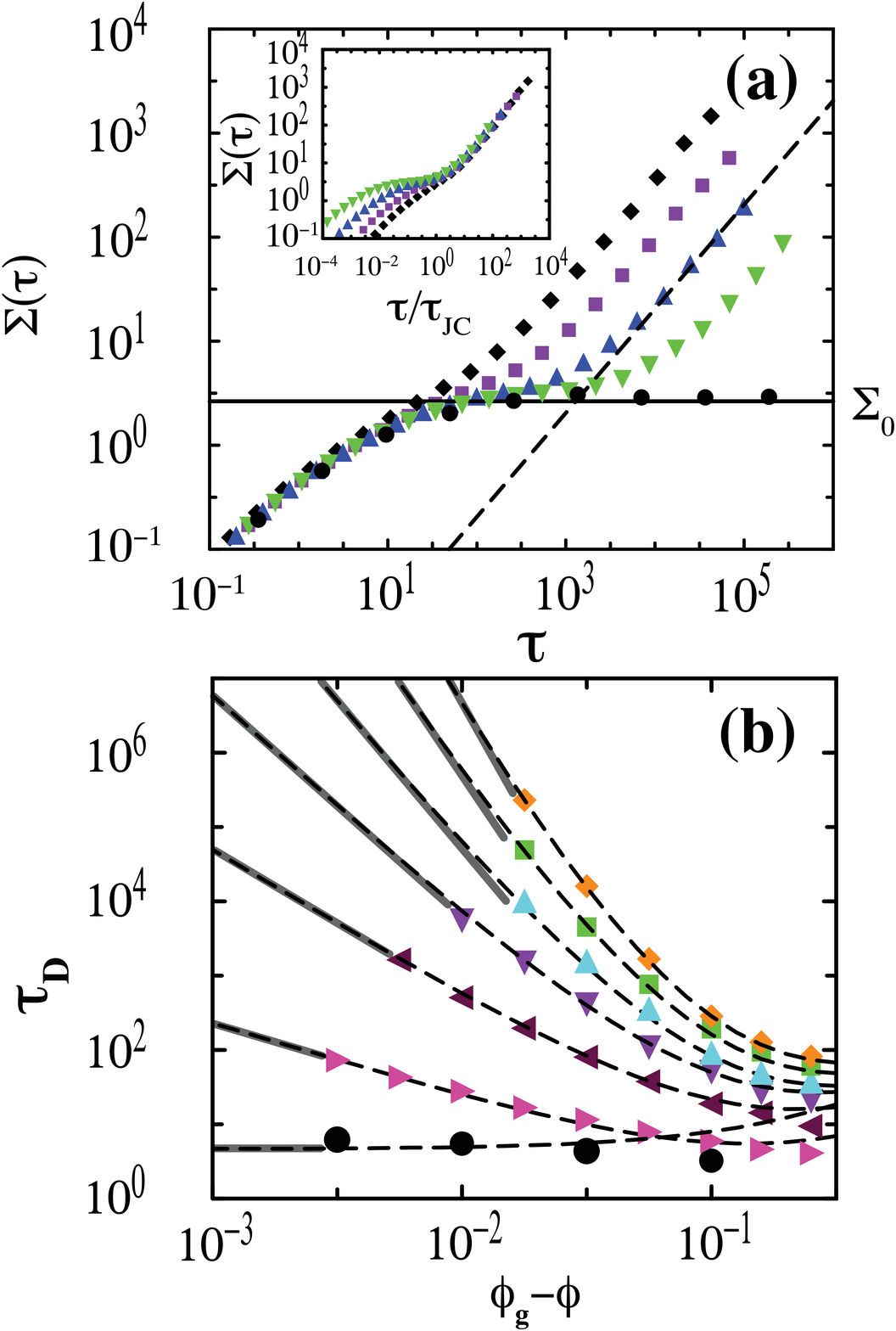}}}
\end{tabular}
\end{center}
\vspace{-0.3in}
\caption{\label{MSDn10} (a) MSD $\Sigma(\tau)$ normalized by
$\Delta^{2}$ for $N=10$ at $\phi_{g}-\phi=0.1$ (diamonds), $0.056$
(squares), $0.032$ (upward triangles), $0.018$ (downward triangles),
and $10^{-5}$ (circles). The time $\tau$ is normalized by
$\Delta^{2}/2D_{s}$, where $D_{s}$ is the short-time diffusion
constant.  The dashed line is a fit to $\Sigma(\tau)=2D\tau$ for large
$\tau$ at $\phi_g-\phi=0.032$.  The inset shows the same data versus
time normalized by the average junction-crossing time $\tau_{JC}$. (b)
Diffusion time $\tau_{D}$ vs. $\phi_g-\phi$ for $N=2,4,6,8,10,12,14$
(from below).  Fits to Eq.~\eqref{diffusion time in terms with scaling
of residence time} with $\alpha$ only weakly depending on $N$ (dashed
lines) and asymptotic behavior \eqref{scaling for time between
switches} (solid gray lines) are also shown.}
\vspace{-0.2in}
\end{figure}

The system dynamics is monitored by measuring the MSD of the blocks,
$\Sigma(\tau) = \sum_{i=1}^{N} \langle[x_i(\tau) - x_i(0)]^2
\rangle/N$, where $x_i(\tau)$ is the position of the center of block
$i$ at time $\tau$.  As seen from the numerical results shown in
Fig.~\ref{MSDn10} (a), for sufficiently large $\phi$ the MSD develops
a plateau $\Sigma=\Sigma_0$, which signals the onset of slow dynamics.
The length of the plateau increases as $\phi$ approaches a critical
value $\phi_g$.  However, for any $\phi < \phi_g$, the MSD finally
becomes diffusive, with $\Sigma(\tau)=2\DlongTime \tau$ as $\tau
\rightarrow \infty$.  This behavior mimics slow dynamics in glassy
materials.  The time $\tau_D = \Sigma_0/2D$ for the system to reach
the long-time diffusive regime is plotted in Fig.~\ref{MSDn10} (b) as
a function of $\phi$ for several system sizes $N$.  The results indicate
that $\tau_D$ exhibits a power-low divergence
\begin{equation}
\label{scaling for time between switches}
\tauDiffusiveStep^{-1}\sim(\phi_g-\phi)^{N/2-1},
\end{equation}
for $\phi \rightarrow \phi_g$ as indicated by the solid lines. 

\begin{figure}
\begin{center}
\begin{tabular}{cccc}
\mbox{\scalebox{0.4}{\includegraphics{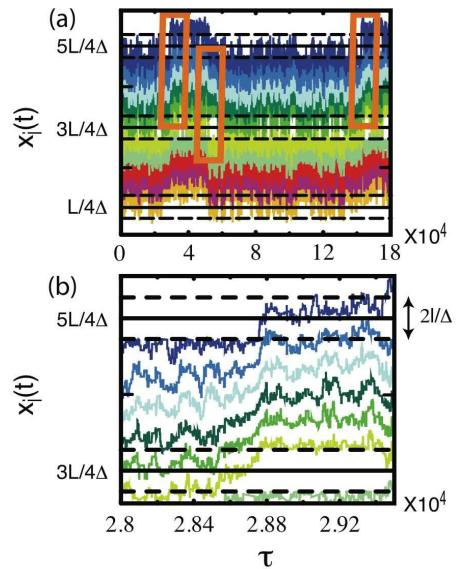}}}
\end{tabular}
\end{center}
\vspace{-0.3in}
\caption{\label{Trajectory} (a) Trajectories $x_i(\tau)$ scaled by
$\Delta$ for each particle $i$ over a time during which three
junction-crossing events (highlighted in orange) occur. The solid
horizontal lines correspond to the center of the junction at $x=(L/4,
3L/4, 5L/4,\ldots)$, and the long-dashed horizontal lines are
displaced by a particle size from them.  (b) Close-up of particle
trajectories during the first junction-crossing event.}
\vspace{-0.2in}
\end{figure}

To gain insight into the system dynamics near $\phi_g$ and explain the
critical behavior \eqref{scaling for time between switches}, we
consider sample trajectories depicted in Fig.\ \ref{Trajectory}.  We
see that most often particles are evenly divided between the two
lobes, and the junction is occupied by one or two particles in the
same mode.  Occasionally, the junction becomes unoccupied and the
direction of motion changes.  However, an analysis of the trajectories
shows that not all such switches produce significant particle
displacements.

On rare occasions, a switch of the direction of motion results in a
significant shift of all particles.  Such a junction-crossing
rearrangement requires that a specific sequence of events unfolds as
shown schematically in Fig.~\ref{Figure-8Geometry} and in actual
trajectories in Fig.~\ref{Trajectory}.  First, a given particle
completely crosses the intersection and enters a lobe that already has
$N/2$ particles, thus creating a compressed lobe with $N/2+1$
particles.  Next, another particle leaves the compressed lobe from the
other end (which requires that the intersection is free) and enters
the uncompressed lobe, so that the particles are again evenly
distributed between the two lobes.  Three such junction-crossing
events are highlighted in Fig.\ \ref{Trajectory} (a), and a closeup of
one of them is shown in Fig.\ \ref{Trajectory} (b).

Due to geometrical constraints, junction crossings can only occur when
$N/2+1\le L/2-1$, which yields $\phi_g = N/(N+4)$.  In the limit $\phi
\rightarrow \phi_{g}$, the available space in the compressed lobe goes
to zero, giving rise to the slow dynamics and the resulting plateau of
the MSD depicted in Fig.~\ref{MSDn10}.  In the inset to
Fig.~\ref{MSDn10} (a), we replot the MSD curves with $\tau$ scaled by
the average time between junction-crossing rearrangements $\tau_{JC}$.
The rescaled MSD curves for different $\phi$ collapse at long times,
which confirms that long-time diffusion is controlled by
junction-crossing events.  An analysis of the geometry of the system
indicates that a junction crossing event produces an average particle shift
$\delta=\frac{3}{2}+\Delta-N^{-1}$, and thus, we find that
$2D=\delta^2/\tau_{JC}$.

Since the slow evolution for $\phi \rightarrow \phi_g$ results from an
entropic bottleneck associated with creation of a compressed lobe, we
expect that the scaling behavior of $\tau_D\sim\tau_{JC}$ with $\Delta
\phi$ can be obtained by calculating the corresponding volume in
configuration space.  In equilibrium, the fraction of time
$\residenceTime/\totalTime$ the system spends in the compressed
configuration can be expressed as
\begin{equation}
\label{fraction of time in switching configuration}
\frac{\residenceTime}{\totalTime}
=\frac{\OmegaLow\,\OmegaUp}{\OmegaTot},
\end{equation}
where $\Omega_{\textit{c,u}}$ and $\OmegaTot$ are configuration
integrals for the compressed (uncompressed) lobe and whole system
respectively.  $\Omega_{c,u}$ and $\OmegaTot$ can be written in terms
of the configurational integral $\OmegaHardRods(L_0,M)=(L_0-M)^M/M!$
for a 1D gas of $M$ unit-length hard rods confined within length $L_0$
(Tonks gas \cite{tonks}).  The compressed and uncompressed lobes
correspond to a Tonks gas of length $\Llobe=L/2-1$ with
$N_{\textit{c,u}}=N/2\pm1$ particles.  Accordingly we have
\refstepcounter{equation}
\label{configuration integrals for lobes}
\begin{equation}
\label{configuration integrals for lobes, a b}
\tag{\theequation{a,b}}
\OmegaLow=\frac{\freeSpaceLow^{N/2+1}}{(N/2+1)!},
\qquad
\OmegaUp=\frac{\freeSpaceUp^{N/2-1}}{(N/2-1)!},
\end{equation}
where $\freeSpaceLow=\halff(L-N-4)$ and $\freeSpaceUp=\halff(L-N)$
denote the free space in the compressed and uncompressed regions.  The
configurational integral for the whole system can be expressed as a
combination of Tonks-gas results,
\begin{equation}
\label{Omega total take 6}
\OmegaTot=2\OmegaHardRods(L,N)
   -\sum_{\stackrel{\scriptstyle N_1,N_2}{N_1+N_2=N}}
    \OmegaHardRods(L_1,N_1)\OmegaHardRods(L_1,N_2),
\end{equation}
where the factor of 2 corresponds to the two directions of motion for
particles in the intersection, and the subtracted sum prevents double
counting configurations with an empty intersection.

At $\phi_g$ the free space in the compressed lobe $\freeSpaceLow$
vanishes, which is the source of the kinetic arrest.  The structural
relaxation time for $\phi\approx\phi_g$ is calculated using
\eqref{fraction of time in switching configuration} to estimate the
average time between junction crossings $\tau_{JC}=T/\numberOfJumps$
(where $\numberOfJumps$ is the number of crossing events).  Assuming
that on average during a single junction-crossing event the system
spends time $\switchTime=\residenceTime/\numberOfJumps$ with one
compressed and uncompressed lobe, we obtain
\begin{equation}
\label{diffusion time in terms of configuration volume}
\tau_D^{-1} \sim \tau_{JC}^{-1} = \frac{\residenceTime}{T}\switchTime^{-1}
  =\frac{\OmegaLow\OmegaUp}{\OmegaTot}\switchTime^{-1},
\end{equation}
which links the structural relaxation time to the ratio of the
configurational integrals \eqref{configuration integrals for lobes}
and \eqref{Omega total take 6}. Using (\ref{configuration integrals
for lobes}a), free space in the compressed lobe can be written in
terms of $\Delta \phi$,
\begin{equation}
\label{scaling for Omega compressed}
\OmegaLow\sim(\phi_g-\phi)^{N/2+1}.
\end{equation}
Thus, one might expect that \mbox{
$\tauDiffusiveStep^{-1}\sim\OmegaLow\sim(\phi_g-\phi)^{N/2+1}$}, i.e.\
the inverse diffusion time scales with the number of transition
configurations.  (This assumption is frequently adopted in analyses of
cooperative glassy dynamics \cite{Gibbs-Adams}).  However, our
numerical results do not support this hypothesis, and instead we
observe a weaker singularity \eqref{scaling for time between
switches}.

The anomalous behavior \eqref{scaling for time between switches} stems
from the fact that not only $\OmegaLow$ but also $\switchTime$ vanishes
at $\phi_g$.  This can be demonstrated by noting that Brownian
dynamics of a 1D gas of hard rods can be mapped onto a 1D system of
point particles with interparticle distances equal to the gaps $\Delta
x_i=x_i-x_{i-1}-1$ between rods in the original Tonks gas.  Since a
gas of point particles does not involve a characteristic lengthscale,
the entire stochastic process $(\Delta x_1(t),\ldots,\Delta x_M(t))$
for systems of different packing fractions (but corresponding initial
conditions) can be scaled onto each other by introducing rescaled
variables
\begin{equation}
\label{rescaled stochastic process}
\Delta x_i(\tilde t)=\lambda^{-1}\Delta x_i(\lambda^{-2}t),
\end{equation}
where $\lambda$ is an appropriate scaling factor (e.g.\ the average
interparticle gap).  In the above relations $\Delta x_1$ is the gap
between the first particle and the position at which the boundary
condition is applied; the boundary condition on the other end of the
domain is represented in terms of the variable $\sum_{i=1}^M\Delta
x_i$.

To apply scaling relation \eqref{rescaled stochastic process} to our
system we recall that a junction-crossing event requires that the
particles are divided between the compressed and uncompressed lobes.
Particles in the compressed lobe evolve as a 1D Tonks gas until an
interaction occurs with a particle that initially resided in the
uncompressed lobe.  We note that the evolution of particles in the
compressed lobe can be rescaled exactly even if a particle leaves this
lobe and enters the junction, because both the particle positions
and boundary conditions can be rescaled.  This is important because a
particle on the border of a junction enters and leaves the
compressed lobe multiple times before a junction-crossing event is
completed.

The mapping \eqref{rescaled stochastic process} of the dynamics of the
compressed lobe implies that the corresponding scaling will also hold
for the average time $\switchTime$ that the system spends in the
compressed configuration.  Taking $\lambda=\freeSpaceLow$, we find
that
\begin{equation}
\label{scaling for residence time}
\switchTime=\alpha\freeSpaceLow^{-2},
\end{equation}
where $\alpha$ is a proportionality constant.  Combining the above
relation with \eqref{diffusion time in terms of configuration volume}
yields~\cite{vogel}
\begin{equation}
\label{diffusion time in terms with scaling of residence time}
\tauDiffusiveStep^{-1}
  =\frac{\alpha\freeSpaceLow^{-2}\OmegaLow\OmegaUp}{\OmegaTot},
\end{equation}
which, according to Eqs.\ \eqref{scaling for Omega compressed} and
\eqref{scaling for residence time}, agrees with the observed anomalous
scaling behavior \eqref{scaling for time between switches}.  In
Fig.~\ref{MSDn10} (b), we show that Eq.\ \eqref{diffusion time in
terms with scaling of residence time} accurately represents the
long-time diffusive dynamics not only in the scaling regime $\phi
\rightarrow \phi_g$, but also at moderate $\phi$.

To summarize, we introduced the Figure-8 model, which exhibits kinetic
arrest when $\phi \rightarrow \phi_g$.  We showed that for
$\phi\approx\phi_g$ long-time diffusion in our system is controlled by
rare, cooperative junction-crossing events, and we determined the
configuration-space volume $\OmegaLow$ corresponding to the transition
states associated with junction crossings.  We also demonstrated that
the inverse structural relaxation time $\tau_D^{-1}$ does not scale
with the volume $\OmegaLow$ as $\phi \rightarrow \phi_g$, but the
scaling also involves an additional singular factor associated with
the accelerated evolution of compressed particle configurations in the
volume $\OmegaLow$.  We predict that similar anomalous behavior may
occur in glassy materials when a cage rearrangement requires
compression of the material surrounding the cage.

There are several possible extensions of the Figure-8 model that may
shed light on important features of the glass transition (such as
dynamic heterogeneities and aging phenomena \cite{cianci,makse}).  We
expect that these phenomena can be characterized by expanding the
approach used here \cite{one}. One
generalization of the Figure-8 model we are now pursuing involves
increasing the number of junctions $j$ and determining how $\phi_g$
depends on $j$ and $N$ in a network of cross-linked channels.  Similar
to the Figure-8 model, the multi-junction system exhibits kinetic
arrest at $\phi_g$ below close packing.  An analysis of this class of
models will thus shed light on the mechanisms that give rise to slow
dynamics in glass-forming materials.

Financial support from NSF grant numbers CBET-0348175 (JB),
CBET-0625149 (PP), and DMR-0448838 (CO) and EPSRC grant number
EP/D050952/1 (RS) is gratefully acknowledged.  We thank G.-J. Gao
for his input and the Aspen Center for Physics where some of this
work was performed.


\begin{references}
\bibitem{vogel} In the large-$N$ limit, $\tau_D \sim
\exp(1/\Delta\phi)$ for $\Delta\phi 
\rightarrow 0$, but this limit is
not the focus of our study.
\end{references}
\end{document}